\documentclass[sigconf]{acmart}
\usepackage{amstext} 
\usepackage{array}  
\usepackage{booktabs} % For formal tables
\usepackage{mathtools} % for "\DeclarePairedDelimiter" macro
\DeclarePairedDelimiter{\floor}{\lfloor}{\rfloor}
\usepackage[colorinlistoftodos]{todonotes}
\usepackage{tablefootnote}
\usepackage{tabularx}

\copyrightyear{2018}
\acmYear{2018}
\setcopyright{acmlicensed}
\acmConference[RecSys Challenge '18]{Proceedings of the ACM Recommender Systems Challenge 2018}{October 2, 2018}{Vancouver, BC, Canada}
\acmBooktitle{Proceedings of the ACM Recommender Systems Challenge 2018 (RecSys Challenge '18), October 2, 2018, Vancouver, BC, Canada}
\acmPrice{15.00}
\acmDOI{10.1145/3267471.3267475}
\acmISBN{978-1-4503-6586-4/18/10}

\usepackage{graphicx}

\begin{document}
\title[Artist-driven layering and user's behaviour impact on recommendations]{Artist-driven layering and user's behaviour impact on recommendations in a playlist continuation scenario}

\author{Sebastiano Antenucci}
\affiliation{%
  \institution{Politecnico di Milano}
}
\email{sebastiano.antenucci@mail.polimi.it}

\author{Simone Boglio}
\affiliation{%
  \institution{Politecnico di Milano}
}
\email{simone.boglio@mail.polimi.it}

\author{Emanuele Chioso}
\affiliation{%
  \institution{Politecnico di Milano}
}
\email{emanuele.chioso@mail.polimi.it}

\author{Ervin Dervishaj}
\affiliation{%
  \institution{Politecnico di Milano}
}
\email{ervin.dervishaj@mail.polimi.it}

\author{Shuwen Kang}
\affiliation{%
  \institution{Politecnico di Milano}
}
\email{shuwen.kang@mail.polimi.it}

\author{Tommaso Scarlatti}
\affiliation{%
  \institution{Politecnico di Milano}
}
\email{tommaso.scarlatti@mail.polimi.it}

\author{Maurizio Ferrari Dacrema}
\orcid{0000-0001-7103-2788}
\affiliation{%
  \institution{Politecnico di Milano}
}
\email{maurizio.ferrari@polimi.it}

\renewcommand{\shortauthors}{Ferrari D. et al.}

\begin{abstract}
In this paper we provide an overview of the approach we used as team Creamy Fireflies for the ACM RecSys Challenge 2018. The competition, organized by Spotify, focuses on the problem of playlist continuation, that is suggesting which tracks the user may add to an existing playlist. The challenge addresses this issue in many use cases, from playlist cold start to playlists already composed by up to a hundred tracks.
Our team proposes a solution based on a few well known models both content based and collaborative, whose predictions are aggregated via an ensembling step.
Moreover by analyzing the underlying structure of the data, we propose a series of boosts to be applied on top of the final predictions and improve the recommendation quality.
The proposed approach leverages well-known algorithms and is able to offer a high recommendation quality while requiring a limited amount of computational resources.
\end{abstract}

\begin{CCSXML}
<ccs2012>
 <concept>
  <concept_desc>Information systems~Recommender systems</concept_desc>
 </concept>
</ccs2012>
\end{CCSXML}

\ccsdesc[100]{Information systems~Recommender systems}

\keywords{ACM RecSys Challenge 2018, Recommendation Systems, Music recommendation, Cold-Start recommendations, Collaborative Filtering}

\maketitle

\section{Introduction}

Recommender systems are a useful tool to offer personalized and relevant content to users in many different sectors like e-commerce or entertainment, in such a way to help the user in identifying relevant content in a wide database. The ACM RecSys Challenge 2018 organized by Spotify focuses on automatic playlist continuation. This domain, as described in \citep{schedl2018current}, is characterized by two important issues that often arise in recommender systems: sparsity of the user-item interactions and the cold-start problem \cite{kaminskas2012contextual}.

Collaborative Filtering (CF) \cite{koren2015advances} is one of the most successful and effective techniques available in recommender systems, however they are prone to rapidly loose their effectiveness when the user-item interactions are sparse. User-based CF considers users to be similar if they tend to interact with items in a similar way, while item-based CF considers tracks to be similar if many users interacted with them in a similar way. With increasing sparsity in the interactions, the ability of CF to accurately inference the similarity between playlists and tracks decreases.

Cold-start problem refers to the task of recommending items to new users and/or recommending new items to users. In case of an almost empty playlist, recommending tracks under the CF framework becomes difficult because there is not enough listening history to make robust recommendations. Also if a new track is added to the system and no user has previously listened to it, it is impossible to find other similar tracks.

In both of these cases Content-Based recommender systems alleviate the problem of recommendation by constructing item-item and user-user similarities from the features available for items and users, respectively \cite{aggarwal2016recommender}. Our team proposes a hybrid recommender system solution to the RecSys Challenge 2018 which merges collaborative filtering and content based techniques while leveraging at the same time both given playlists' structure and domain knowledge. As per competition rules, the source code is publicly available \footnote{https://github.com/MaurizioFD/spotify-recsys-challenge}.

The rest of the paper is organized as follows. In Section \ref{sec:problem_formulation} we outline the problem formulation, the dataset structure and the evaluation metrics. In Section \ref{sec:preprocessing} we describe the preprocessing steps. In Section \ref{sec:algorithms} we list the algorithms we have used. In Section \ref{sec:ensemble} we describe the ensemble structure and in Section \ref{sec:postprocessing} the post-processing steps and boosts. Section \ref{sec:creative_track} describes how we addressed the creative track and which data we used. Finally Section \ref{sec:computational_requirements} lists the computational requirements for the various phases of our proposed solution.

\section{PROBLEM FORMULATION}
\label{sec:problem_formulation}
The RecSys Challenge focuses on the music recommendation task, in particular on automatic playlist continuation. The goal is to develop a recommender system which is able, given a playlist and some related information to generate a list of recommended tracks that can be added to that playlist, thereby "continuing" it. The challenge is split in two parallel tracks with different rules:

\begin{itemize}
\item \textit{Main track}: only the \textit{Million Playlist Dataset} (MPD) \footnote{https://recsys-challenge.spotify.com} can be used to train the recommender system.
\item \textit{Creative track}: external, public and freely available data sources are allowed in order to enrich the \textit{MPD} and improve the quality of the recommendations.
\end{itemize}

\subsection{Dataset description}
Spotify provided for the competition two different datasets:

\begin{itemize}
\item \textit{The Million Playlist Dataset:} contains $1M$ playlists created by users on the Spotify platform. These playlists were created between January 2010 and October 2017. Each playlist contains a title, the track list (including track metadata) editing information (last edit time, number of playlist edits) and other miscellaneous information about the playlist.

\item \textit{The Challenge Set:} contains $10K$ incomplete playlists. The challenge is to recommend 500 tracks for each of these playlists. Playlists are grouped into 10 different categories, with 1K playlists in each category:
\newline
\begin{enumerate}
\item Playlists with title only
\item Playlists with title and the first track
\item Playlists with title and the first 5 tracks
\item Playlists with first 5 tracks (no title)
\item Playlists with title and the first 10 tracks
\item Playlists with first ten tracks (no title)
\item Playlists with title and the first 25 tracks
\item Playlists with title and 25 random tracks
\item Playlists with title and the first 100 tracks
\item Playlists with title and 100 random tracks
\end{enumerate}
    
\end{itemize}

For further details of the two datasets refer to the dataset website.

\subsection{Evaluation metrics}
Predictions are evaluated according to three different metrics and final rankings are computed using the Borda Count election strategy.
Assuming the ground truth set of tracks defined by $G_{t}$, and the ordered list of recommended tracks by $R_{t}$:

\begin{itemize}
\item \textbf{R-precision:} this metric is evaluated both at the track level (e.g., tracks correctly recommended) and at the artist level (e.g., any other track by the same artist). The track level is computed as follows:
\[
    Rprec_{t} = \frac{| G_{t} \cap R_{t_{1:|G_{t}|}}|}{|G_{t}|}
\]

Being $G_{a}$ the ground truth set of unique artists of $G_{t}$ and $R_{a}$ is the ordered list of recommended artists of $R_{t}$ for all the tracks which have not been matched at the track level, the artist level is computed as follows:
\[
    Rprec_{a} = \frac{| G_{a} \cap R_{a} |}{|G_{a}|}
\]
A match at the artist level can only be counted once per artist per playlist.
%i.e. after the first artist match, subsequent matches to that same artist do not contribute any additional score.
The final score is:
\[
    Rprec = Rprec_{t} + 0.25 * Rprec_{a}
\]
\item \textbf{NDCG:} normalized discounted cumulative gain is a well known rank-based metric used in recommender system.
%(DCG) measures the ranking quality of the recommended tracks, increasing when relevant tracks are placed higher in the list. Normalized DCG (NDCG) is determined by calculating the DCG and dividing it by the ideal DCG in which the recommended tracks are perfectly ranked.

% \[
%     DCG = rel_{1} + \sum_{i=2}^{|R_{t}|} \frac{rel_{i}}{log_{2}(i+1)}
% \]
% The IDCG is equal to:
% \[
%     IDCG = 1 + \sum_{i=2}^{|G_{t}|} \frac{1}{log_{2}(i+1)}
% \]
% If the size of the set intersection of 
% $G_{t}$ and $R_{t}$ is empty, then the DCG is equal to 0. The NDCG metric is calculated as:
% \[
%     NDCG = \frac{DCG}{IDCG}
% \]

\item \textbf{Recommended Songs clicks:} is computed as follows:
%\vspace{-2 mm}
\[
clicks = \floor*{\frac{argmin_{i} 
\left\{ 
   R_{t_i}: R_{t_i} \in G_{t} 
\right\} - 1 }{10}}
\]
\vspace{0 mm}
where $R_{t_i}$ is the track that occupies the $i$th index of the ordered list of recommended tracks $R_{t}$.
%It is based on the homonym Spotify feature that, given a set of tracks in a playlist, recommends 10 tracks to add to the playlist. 
Recommended Songs clicks is the number of refreshes needed before a relevant track is encountered.

\end{itemize}

% \subsection{Local validation}
% \mf{maybe not relevant}
% In order to locally test our algorithms, we split the dataset into training and validation set. We observe the statistical properties of the \textit{Challenge Set} in order to create a validation set as similar as possible.
% For each category of the \textit{Challenge Set} we take into account the maximum and the minimum number of holdouts of the playlists, which is the number of tracks that have been removed. We sample randomly $1K$ playlists from the \textit{Million Playlists Dataset} which have a number of holdouts strictly included in the aforementioned range.    

\section{PREPROCESSING}
\label{sec:preprocessing}

In order to address the cold-start problem in first category, where we have no available interactions for playlists, we apply information retrieval techniques to build a feature space from playlists titles. The following preprocessing steps have been adopted: 
\begin{enumerate}
\item Removing spaces from titles made by only separated single letters. 
\item Elimination of uncommon characters like dots and brackets.
\item Separation of words composed by letters and numbers, appending the resulting new tokens. 
\item Appending to the title the Lancaster and Porter stemming of title's words.
\end{enumerate}
The result is a list that contains all the original tokens and the newly created ones.

\subsection{Track position and Artist Heterogeneity}
\label{sec:ArH}
\begin{figure}[t]
  \includegraphics[width=0.45\textwidth]{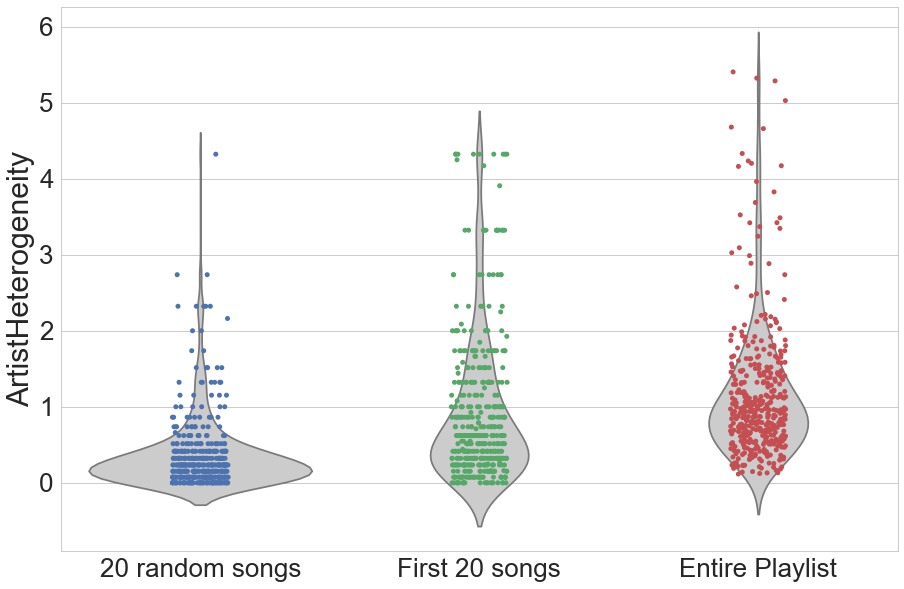}
  \caption{ArtistHeretogeneity for 1000 long playlists. The gray area behind the points shows the Distribution of ArH, over three different sampling strategies of songs in a long playlist.}
  \label{fig:ArH}
\end{figure}

In this music recommendation domain playlists are created by users and sometimes exhibit a common underlying structure due to the way a user fills them. 
\begin{enumerate}
\item Adding all the songs from the same Album one after another.
\item Creating playlists with tracks from only one Artist and the featurings that involve him. 
\item Creating a long playlist of one genre and fill them in the years with new songs of the same artist.
\item Creating a playlist with many different artists in the first tracks, and readding the same artists later on as seen in Figure \ref{fig:ArH}.
\end{enumerate}

To leverage these patterns we define a new measure to estimate how diverse the artists are.
\vspace{-3 mm}
\[
ArtistHeterogeneity_{p} = \log _{2} \left (    \frac{ \ \left | \  uniqueTracks_{p} \   \right  | \ }{ \ \left  | \ uniqueArtists _{p} \  \right  |  \ }     \right )  
\]
\vspace{0 mm}
Where p is the playlist.
A value of $ArH=0$ refers to a playlist with a lot a different artists, like the ones with the Top100 songs of a genre.
A high value of $ArH$ points to a very repetitive playlist.

In particular, consider a set of 500 long playlists (from 100 to 250 songs). If you observe the first 20 songs or the same number of songs but sampled at random you will obtain very different values for ArtistHeterogeneity, see Figure \ref{fig:ArH}.
We are also able to exploit the behaviour highlighted by this results by applying both boosts and clustering, as it is described in Section \ref{sec:clusterArH} and  \ref{sec:boost}

\section{ALGORITHMS}
\label{sec:algorithms}
Our model is composed by five well known algorithms, some content based, some collaborative and some item-based and some user-based. In the following sections we will call \emph{playlist-track matrix} (PTM) the matrix having the playlists as rows, tracks as columns and a cell value of one if the track belongs to that playlist.

\subsection{Personalized Top Popular}
For each element from category two, we implemented a personalized top popular algorithm based on the only track in the playlist.
\subsubsection{Track Based}
We select all the playlists containing that track and compute the top popular on them.
\subsubsection{Album Based}
Given the track's album, we select all the playlists containing tracks from that album and compute the top popular on them.

\subsection{Collaborative Filtering - Track Based}
In the track-based CF algorithm first we apply BM25 \citep{Robertson:2009:PRF:1704809.1704810} normalization on the PTM, then we define the similarity between two tracks i and j as the dot product between the corresponding PTM columns:
\[
    s_{ij} = r_{i} * r_{j}
\]
The score prediction of target track i for playlist u is given by:
\[
    r_{ui} = \sum_{j \epsilon I(u)}^{KNN} r_{uj} * (s_{ji})^ p
\]
Where KNN are the top k-nearest neighbors and \textit{p} is a coefficient that helps to discriminate values of the similarity.

\subsection{Collaborative Filtering - Playlist Based}
We define the similarity between two playlists i and j as the Tversky  \citep{Tversky1977-TVEFOS}  coefficient between the corresponding PCM rows:
\[
    s_{ij} = \frac{r_{i} * r_{j}}{\alpha (|r_{i}| - r_{i} * r_{j}) + \beta (|r_{j}| - r_{i} * r_{j}) + r_{i} * r_{j} + h}
\]
Where $\alpha$ and $\beta$ are the Tversky coefficients between [0,1] and \textit{h} is the shrink term.
\newline
The score prediction of target item i for user u is given by:
\[
    r_{ui} = \sum_{v \epsilon U(i)}^{KNN} (s_{uv})^ p * r_{vi} 
\]

\subsection{Content Based Filtering - Track Based}
For the track-based content based filtering (CBF) we first applied BM25 normalization on the track-content matrix associating each track to its features, next we define the similarity between two tracks i and j as the dot product between the two feature vectors:
\[
    s_{ij} = f_{i} * f_{j}
\]
The score prediction of target item i for user u is given by:
\[
    r_{ui} = \sum_{j \epsilon I(u)}^{KNN} r_{uj} * (s_{ji})^ p
\]
We run three different types of CBF, each one based on different combination of item features: Artist ID, Album ID, Album ID together with artist ID.

For the last one, after a phase of tuning, we assign different weight in the track content matrix giving more weight to the album features since it provides us better overall score.

\subsection{Content Based Filtering - Playlist Based}
We applied two different playlist-based CBF algorithms.

\subsubsection{Track features} We build a playlist content matrix in which we represent playlists with the feature of the tracks they contain.
%\[
%    UCM = PCM * ICM
%\]
%Where URM is the user-rating matrix and ICM the item-content matrix.

We then again built three different user-content matrix using different combinations of track features: Artist ID, Album ID, Album ID together with artist ID.

%For the last one, after a phase of tuning, we assign different weight in the item content matrix giving more weight to the album features since it provide us better overall score.

Next we apply BM25 on the playlist content matrix and we compute the similarity between two playlists i and j as the Tversky coefficient between the two playlist-feature vectors.

%\[
%    s_{ij} = \frac{f_{i} * f_{j}}{\alpha (|f_{i}| - f_{i} * f_{j}) %+ \beta (|f_{j}| - f_{i} * f_{j}) + f_{i} * f_{j} + h}
%\]

%\newline
%The score prediction of target item i for user u is given by:
%\[
%    r_{ui} = \sum_{v \epsilon U(i)}^{KNN} (s_{uv})^ p * r_{vi} 
%\]

\subsubsection{Playlist name} We create a playlist-content matrix for the tokens extracted from titles. Some playlists have untokenizable titles (e.g., emojis) to avoid empty recommendations and to improve accuracy we ensembled two different approaches relying on the playlist title
\begin{enumerate}
	\item Content based filtering (CBF) based on the tokens extracted from the titles in the preprocessing phase.
    \item Content based filtering (CBF) based on an exact title match.
\end{enumerate}
The final model is a weighted sum of the recommendations of the two aforementioned approaches.

\subsection{Parameters tuning}
For each algorithm and for each category on which we are evaluated, we tune the parameters on our validation set.

\begin{itemize}
\item number of k-nearest neighbors (KNN) of the similarity matrices
\item the power coefficient \textit{p} for the similarity values
\item the coefficients $\alpha$ and $\beta$ of the Tversky similarity
\item the shrink term \textit{h}
\end{itemize}
 
\section{ENSEMBLE}
\label{sec:ensemble}
\subsection{ Base Ensemble}
\begin{figure}[h]
  \includegraphics[width=0.45\textwidth]{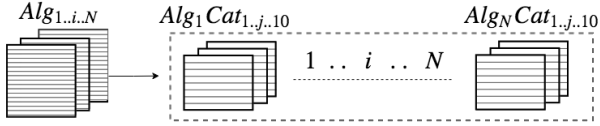}
  \caption{Category splitting of each algorithm.}
  \label{img:cat_div}
\end{figure}

An analysis on the case of study makes us observe that the different algorithms are better suited for subsets of playlists with specific characteristics. Content base approaches fit well on short playlists with similar features, on the other hand, collaborative filtering approaches gave us the best results on long and heterogeneous playlists.\newline
We divide the results by category as it is shown in Figure \ref{img:cat_div}.
If we consider N algorithms, we use them to compute, for each playlist, N sets of tracks scores such that the highest valued tracks will be recommended to that playlist.

The final model is a weighted sum of the N score predictions taking into account the length of the playlist and the position of the tracks. This allows to take advantage of the diversity in the predictions made by the different algorithms.

\subsection{ArH-based Cluster Ensemble}
\label{sec:clusterArH}
One of the characteristics we took into account is the Artist Heterogeneity (ArH). Playlists are assigned to a cluster based on their ArH index, see Table \ref{table:ArH_clusters}. 
\newcolumntype{L}{>{$}l<{$}} 
\begin{table}[h]
\begin{tabular}{L|cccc}
  \hline
   & Cluster1  & Cluster2 & Cluster3 & Cluster4  \\ \hline
   ArH_{p} & =0  & <1 & <2 & >=2  \\ \hline
\end{tabular}
\caption{Artist Heterogeneity clusters}
\label{table:ArH_clusters}
\end{table}

\begin{figure}[t]
  \includegraphics[width=0.45\textwidth]{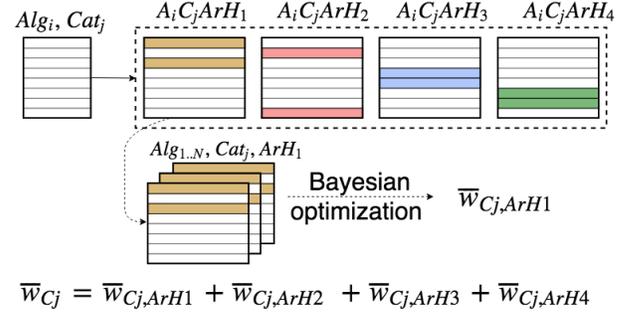}
  \caption{Cluster division and Bayesian optimization}
   \label{img:clus_div}
\end{figure}

\textit{For each category} and \textit{for each cluster} we search the best model weights by bayesian optimization, Figure \ref{img:clus_div}.

This ensemble also takes into account the cluster of the playlist.
For the categories 4,5,6,8,10 this approach gives better results than the base ensemble (see Table \ref{table:clusAH_res}).
The final model for these categories is still a weighted sum of the N score predictions, but taking into account the ArH cluster.

\begin{table}[t]
\begin{tabular}{l|ccc} 
\hline
  	& clicks & NDCG & R-precision \\ \hline
 ArH-based & -0.0322 & +0.0058 & +0.0041 \\ \hline
\end{tabular}
\caption{Relative improvement for categories 4, 5, 6, 8, 10 after playlist's clusterization.}
\label{table:clusAH_res}
\end{table}

\section{POSTPROCESSING}
\label{sec:postprocessing}

Once we apply our \textit{per-category ensemble} technique, we obtain a new set of predictions which takes into account the recommendations of each algorithm. We improve our score leveraging on domain-specific patterns of the dataset.

\subsection{Boosts}
\label{sec:boost}
The following boosts share a common work flow: they start from a list of $K$ predicted tracks for a playlist $p$ and for each $k \in K$ they boost the $Score_{p_k}$ computed via the ensemble model in this way:
\[
Score_{p_k} = Score_{p_k} + Boost_{p_k}
\]

\subsubsection{Gap Boost}
It is an heuristic which applies to playlists of category 8 and 10, where known tracks for each playlist are distributed at random. Since known tracks are not in order, there exist "gaps" between each pair of known tracks. We exploit this information by reordering our final prediction giving more weight to tracks which seems to better "fit" between all the gaps of the playlist.
Therefore, for each playlist $p$ we select the first $K$ tracks of the prediction and we add to each of them the following value:
\[
GapBoost_{p_k} = \gamma\sum_{g \in G} \frac{S_{k, g_l}\: S_{k, g_r}}{d_g}\quad \forall k \in K
\]
where $S$ is a similarity matrix between tracks obtained by a content based filtering recommender as described in section 4.4, $G$ is the set of all the gaps in the playlist, $g_{l}$ and $g_{r}$ are the tracks which correspond respectively to the left boundary and the right boundary of the gap, $d_{g}$ is the length of the gap (the difference of the position in the playlist of the boundary tracks) and $\gamma$ is a weight factor. This technique improves significantly the R-precision and the Recommended clicks metrics, leaving the NDCG almost unchanged. 

\subsubsection{Tail Boost}
We apply this technique to categories 5, 6, 7, 9, where known tracks for each playlist are given in order. The basic idea behind this approach is that the last tracks are the most informative about the "continuation" of a playlist, therefore we boost all the top tracks similar to the last known tracks, starting from the tail and proceeding back to the head with a discount factor.

\subsubsection{Album Boost}
This approach leverages the fact that some playlists are built collecting tracks in order from a specific album. Therefore in categories 3, 4, 7 and 9, where known tracks for each playlist are given in order, we use this heuristic to boost all the tracks from a specific album where the last two known tracks belong to the same album. Album Boost improves the Recommender Songs clicks metric.

\begin{table}[t]
\begin{tabular}{l|ccc} 
\hline
 Algorithm 	& clicks & NDCG & R-precision \\ \hline
 Gap Boost 	& -0.0003 & +0.0001 & +0.0021 \\
 Tail Boost & -0.0060 & +0.0015 & +0.0008 \\ 
 Album Boost & -0.0230 & +0.0011 & +0.0005 \\ \hline
\end{tabular}
\caption{Relative improvement of each boost on the three metrics.}
\label{table:boosts}
\end{table}

\section{Creative Track}
\label{sec:creative_track}
Our approach to the creative track was heavily inspired by the approach used to compete in the main track. The rules of the competition specified that to qualify as successful, the final submission to the creative track must use external sources with the condition of being \textbf{public and freely accessible to all participants}.

\subsection{External Datasets}
Under the rules imposed by the competition organizers, we explored the datasets on table \ref{table:ext_datasets}.

\begin{table}[t]
\begin{tabular}{llcl} 
\hline
 Dataset Name               				& Data Type             & Year  \\ \hline
 \#nowplaying music \tablefootnote{\#nowplaying dataset http://dbis-nowplaying.uibk.ac.at} 			& Listening behavior  	& 2018  \\ 
 \#nowplaying playlists     				& Playlist  			& 2015   \\
 MLHD \tablefootnote{The Music Listening Histories Dataset (MLHD) http://ddmal.music.mcgill.ca/research/musiclisteninghistoriesdataset}          		    		& Listening behavior  	& 2017  \\
 FMA  \tablefootnote{A Dataset For Music Analysis https://arxiv.org/abs/1612.01840}                       	& Audio Features  		& 2017  \\ 
 MSD  \tablefootnote{Million Song Dataset https://labrosa.ee.columbia.edu/millionsong/}                       	& Audio Features 		& 2011  \\ 
 Spotify API \tablefootnote{https://developer.spotify.com/documentation/web-api/} 					& Audio Features, popularity	& 2018 \\ \hline
\end{tabular}
\caption{External datasets explored for the creative track. Listening behaviour refers to timestamps of listening events.}
\label{table:ext_datasets}
\end{table}
We spent considerable effort in trying to reconcile the tracks from the Million Playlist Dataset (MPD) provided by Spotify with those from external datasets but matching the name of the tracks and artists proved to be difficult and error-prone.  
Spotify Web API, on the other hand, being an API provided by Spotify itself, allowed us to retrieve for all tracks in MPD and in the Challenge Dataset the following features (using \verb|audio-features/{id}| and \verb|tracks/{id}| endpoints): acousticnes, danceability, energy, instrumentalness, liveness, loudness, speechiness, tempo, valence, popularity.
During the data collection process, we found 159 tracks with missing audio features. As a preprocessing step, we filled in missing values for 159 tracks with the respective mean over all available data. For those tracks we are not able to retrieve \emph{popularity} feature therefore we considered them as non-popular tracks, and filled in the missing values with 0, the lowest popularity level.
In this way, we obtain a complete enriched dataset which contains 2,262,292 tracks and corresponding audio features and popularity. 

\subsection{Audio Feature Layered Content Based Filtering - Track Based}
Inspired by the content based filtering (CBF) approach in the main track, we implemented a creative CBF which is able to adjust the artist based track recommendation using ten additional features from our enriched dataset. 

In order to better illustrate the idea, we give a graphical representation of the item content matrix (ICM) by random sampling 200 artists.% As shown in Figure \ref{fig:icm}. 
The track-track similarity matrix calculated with a normal CBF, as used in the main track, is not able to distinguish tracks belonging to the same artist. We call it \textbf{artist-level similarity}. This is clear if we take into consideration row $i$ of the ICM which represents a track; in $i$ only one column has a value of 1 corresponding to the artist of the track. For two tracks to be similar they must share the same artist thus making the similarity values of all $K$ most similar tracks to $i$ equal. In this way we cannot differentiate between similar tracks of \emph{i}.

\begin{figure}[t]
  \includegraphics[width=0.45\textwidth]{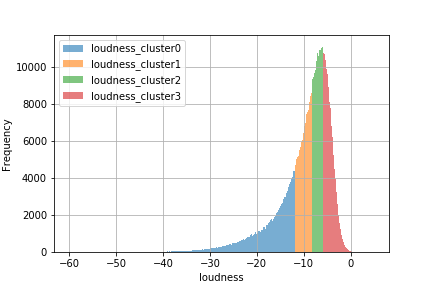}
  \caption{Clustering of tracks over \emph{loudness} feature. The colors represent the different clusters.}
  \label{fig:cluster_loudness}
\end{figure}

\begin{figure}[t]
  \includegraphics[width=0.45\textwidth]{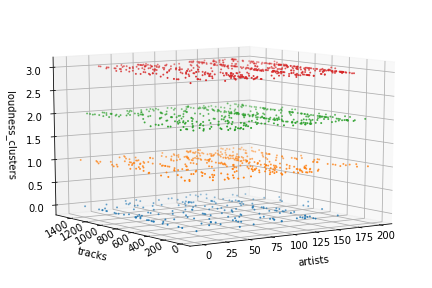}
  \caption{Layered ICM over \emph{loudness} feature (200 sampled artists).}
  \label{fig:loudness_layered_icm}
\end{figure}

The creative CBF is implemented with the following steps:
\begin{enumerate}
\item Divide the tracks into 4 clusters with equal number of elements, according to each feature. Take the \emph{loudness} feature as an example, the clustering result is shown in Figure \ref{fig:cluster_loudness}.
\item Considering feature clusters as a 3rd dimension, split the dense ICM into 4 sparse layers. A \emph{loudness} based layered ICM is illustrated in Figure \ref{fig:loudness_layered_icm}.% For comparison, the sparsity of layer0 is shown in Figure \ref{fig:icm_layer0}. 
\item Concatenate 4 layers of sparse matrices horizontally in order to create a final sparsified ICM. 
\item Applying the CBF approach to the sparsified ICM, we can calculate a \textbf{sub-artist-level} track-track similarity.
\end{enumerate}
In practice, this creative CBF is able to improve the artist-based recommendations in all three evaluation metrics. The comparison is presented on table \ref{table:cbfi_ev_tr} and \ref{table:cbfi_ev_ar}.

\begin{table}[t]
\begin{tabular}{l|ccc} 
\hline
 Algorithm 	 	& clicks 		& NDCG 		  & R-precision 	\\ \hline
 CBF 			& 19.9644       & 0.075191    & 0.037054  		 \\
 creative CBF	& 17.1286       & 0.086872    & 0.050467 		 \\ \hline
\end{tabular}
\caption{Track level local evaluation of artist based CBF and creative CBF (mean value of 10 categories)}
\label{table:cbfi_ev_tr}
\end{table}

\begin{table}[t]
\begin{tabular}{l|ccc} 
\hline
 Algorithm 	 	& clicks 		 & NDCG 		& R-precision			\\ \hline
 CBF 			& 17.4876        & 0.123445     & 0.003055          \\
 creative CBF	& 13.2587        & 0.159920     & 0.006016          \\ \hline
\end{tabular}
\caption{Artist level local evaluation of artist based CBF and creative CBF (mean value of 10 categories)} 
\label{table:cbfi_ev_ar}
\end{table}

% \begin{figure}[t]
%   \includegraphics[width=0.45\textwidth]{img/icm.png}
%   \caption{ICM (200 sampled artists). The x-axis represents the ID of 200 unique artists, while the y-axis stands for unique track IDs. The black dot in each row indicates that this track belongs to the corresponding artist.}
%   \label{fig:icm}
% \end{figure}

% \begin{figure}
%   \includegraphics[width=0.45\textwidth]{img/icm_layer0.png}
%   \caption{\emph{loudness} sparsified ICM layer0 (200 sampled artists)\mf{maybe remove}}
%   \label{fig:icm_layer0}
% \end{figure}

\subsection{Feature Layered Collaborative Filtering - Track Based}
Following the sparsifying idea in the previous subsection, we implement a layering procedure also to the playlist-track matrix. In creative track, the track features we used for layering procedure are: all feature clusters, album, artist. While in the main track, the layering idea is applied with only album and artist feature.
Due to the different character of each category of playlists, in practice, we found out that this sparsified PTM has a good effect on the recommendation of the eighth and tenth category of playlists.

\section{COMPUTATIONAL REQUIREMENTS}
\label{sec:computational_requirements}
To run the entire model we use a AWS memory optimized cr1.8xlarge VM with 32 vCPU and 244 GiB of RAM.

\begin{table}[t]
\begin{tabular}{l|llc} 
\hline
Step						& Time 		& RAM 			& Model dependent	\\  \hline
Fast Models Creation \ref{sec:algorithms}& 1h  		& 150GB 		& Yes 		\\ 
Normal Models Creation \ref{sec:algorithms}  	& 1.5h  	& 80GB 			& Yes 		\\ 
Bayesian Optimization \ref{sec:clusterArH}	& 16h 		& ~15GB			& No 		\\ 
Ensemble \ref{sec:ensemble}			& 5m  		& <8GB 			& Yes 		\\ 
Postprocessing \ref{sec:boost}  			& 8m  		& <8GB 			& Yes 		\\ 
\hline
\end{tabular}
\caption{Computational requirements for the different steps. Model  dependent steps need to be done again to recommend new playlists}\end{table} 

The model creation can be done in two different ways depending on the needs and available resources.\newline
The search of best parameters takes up to 16 hours on the appointed machine but is a procedure that needs to be computed only one time.

\section{RESULTS AND CONCLUSION}
Our recommendation architectures allowed us to reach the 4th place in the main track and the 2nd place in the creative track.
The scores of our final model on both main and creative track are reported in Table \ref{table:final_leaderboard}. These scores are evaluated against 50 \% of the challenge set, as stated in the website of the challenge. The predictions of most of the algorithms in the ensemble are heavily correlated. Nevertheless, the ensemble manages to extract the differing predictions from each algorithm, which is beneficial for the evaluation score.
The major strength of our architecture is that is built in a simple and modular way. It can be easily extended with additional features coming from different datasets and new techniques can be implemented with no impact on the pre-existent work flow. Furthermore our architecture relies on an efficient Cython implementation of the most computationally intensive tasks, which allows to keep the time and space complexity under a reasonable threshold.

\begin{table}[H]
\begin{tabular}{l|rrr} 
\hline
 Track 	& clicks & NDCG & R-precision \\ \hline
 Main 	& 1.9810 & 0.3867 & 0.2207 \\
 Creative & 1.9596 & 0.3858 & 0.2206 \\ \hline
\end{tabular}
\caption{Public Leaderboard scores of our final model on both main and creative track.}
\label{table:final_leaderboard}
\end{table}

\section*{Acknowledgement}
The authors would like to thank Prof. Paolo Cremonesi for his continuous support and for inspiring us to choose the team's name.

\bibliographystyle{ACM-Reference-Format}
\bibliography{references}


\begin{thebibliography}{6}

%%% the Web address.

\ifx \showCODEN    \undefined \def \showCODEN     #1{\unskip}     \fi
\ifx \showDOI      \undefined \def \showDOI       #1{#1}\fi
\ifx \showISBNx    \undefined \def \showISBNx     #1{\unskip}     \fi
\ifx \showISBNxiii \undefined \def \showISBNxiii  #1{\unskip}     \fi
\ifx \showISSN     \undefined \def \showISSN      #1{\unskip}     \fi
\ifx \showLCCN     \undefined \def \showLCCN      #1{\unskip}     \fi
\ifx \shownote     \undefined \def \shownote      #1{#1}          \fi
\ifx \showarticletitle \undefined \def \showarticletitle #1{#1}   \fi
\ifx \showURL      \undefined \def \showURL       {\relax}        \fi
% invisible to TeX
\providecommand\bibfield[2]{#2}
\providecommand\bibinfo[2]{#2}
\providecommand\natexlab[1]{#1}
\providecommand\showeprint[2][]{arXiv:#2}

\bibitem[\protect\citeauthoryear{Aggarwal et~al\mbox{.}}{Aggarwal
  et~al\mbox{.}}{2016}]%
        {aggarwal2016recommender}
\bibfield{author}{\bibinfo{person}{Charu~C Aggarwal} {et~al\mbox{.}}}
  \bibinfo{year}{2016}\natexlab{}.
\newblock \bibinfo{booktitle}{\emph{Recommender systems}}.
\newblock \bibinfo{publisher}{Springer}.
\newblock


\bibitem[\protect\citeauthoryear{Kaminskas and Ricci}{Kaminskas and
  Ricci}{2012}]%
        {kaminskas2012contextual}
\bibfield{author}{\bibinfo{person}{Marius Kaminskas} {and}
  \bibinfo{person}{Francesco Ricci}.} \bibinfo{year}{2012}\natexlab{}.
\newblock \showarticletitle{Contextual music information retrieval and
  recommendation: State of the art and challenges}.
\newblock \bibinfo{journal}{\emph{Computer Science Review}}
  \bibinfo{volume}{6}, \bibinfo{number}{2-3} (\bibinfo{year}{2012}),
  \bibinfo{pages}{89--119}.
\newblock


\bibitem[\protect\citeauthoryear{Koren and Bell}{Koren and Bell}{2015}]%
        {koren2015advances}
\bibfield{author}{\bibinfo{person}{Yehuda Koren} {and} \bibinfo{person}{Robert
  Bell}.} \bibinfo{year}{2015}\natexlab{}.
\newblock \showarticletitle{Advances in collaborative filtering}.
\newblock In \bibinfo{booktitle}{\emph{Recommender systems handbook}}.
  \bibinfo{publisher}{Springer}, \bibinfo{pages}{77--118}.
\newblock


\bibitem[\protect\citeauthoryear{Robertson and Zaragoza}{Robertson and
  Zaragoza}{2009}]%
        {Robertson:2009:PRF:1704809.1704810}
\bibfield{author}{\bibinfo{person}{Stephen Robertson} {and}
  \bibinfo{person}{Hugo Zaragoza}.} \bibinfo{year}{2009}\natexlab{}.
\newblock \showarticletitle{The Probabilistic Relevance Framework: BM25 and
  Beyond}.
\newblock \bibinfo{journal}{\emph{Found. Trends Inf. Retr.}}
  \bibinfo{volume}{3}, \bibinfo{number}{4} (\bibinfo{date}{April}
  \bibinfo{year}{2009}), \bibinfo{pages}{333--389}.
\newblock
\showISSN{1554-0669}
\urldef\tempurl%
\url{https://doi.org/10.1561/1500000019}
\showDOI{\tempurl}


\bibitem[\protect\citeauthoryear{Schedl, Zamani, Chen, Deldjoo, and
  Elahi}{Schedl et~al\mbox{.}}{2018}]%
        {schedl2018current}
\bibfield{author}{\bibinfo{person}{Markus Schedl}, \bibinfo{person}{Hamed
  Zamani}, \bibinfo{person}{Ching-Wei Chen}, \bibinfo{person}{Yashar Deldjoo},
  {and} \bibinfo{person}{Mehdi Elahi}.} \bibinfo{year}{2018}\natexlab{}.
\newblock \showarticletitle{Current challenges and visions in music recommender
  systems research}.
\newblock \bibinfo{journal}{\emph{International Journal of Multimedia
  Information Retrieval}} \bibinfo{volume}{7}, \bibinfo{number}{2}
  (\bibinfo{year}{2018}), \bibinfo{pages}{95--116}.
\newblock


\bibitem[\protect\citeauthoryear{Tversky}{Tversky}{1977}]%
        {Tversky1977-TVEFOS}
\bibfield{author}{\bibinfo{person}{Amos Tversky}.}
  \bibinfo{year}{1977}\natexlab{}.
\newblock \showarticletitle{Features of Similarity}.
\newblock \bibinfo{journal}{\emph{Psychological Review}} \bibinfo{volume}{84},
  \bibinfo{number}{4} (\bibinfo{year}{1977}), \bibinfo{pages}{327--352}.
\newblock


\end{thebibliography}

\end{document}